\documentclass[12pt]{article}
\usepackage{bbm}
\usepackage{amsfonts}
\usepackage{amsmath}
\usepackage{latexsym,cite,amssymb}
\usepackage{times}
\usepackage{graphicx}
\usepackage{color}
\usepackage{yfonts}
\usepackage{bm}
\def\be{\begin{equation}}
\def\ba{\begin{eqnarray}}
\def\ee{\end{equation}}
\def\ea{\end{eqnarray}}
\def\no{\nonumber\\}
\def\wn{\textswab{w}}
\def\qn{\textswab{q}}
\def\deriv{\partial}

\def\calA{{\cal A}}\def\calB{{\cal B}}

\makeatletter
\renewcommand{\theequation}{\thesection.\arabic{equation}}
\@addtoreset{equation}{section} \makeatother
\input amssym.def
\input amssym.tex

\begin{document}

\title{
Hydrodynamics of  RN AdS$_4$ black hole and  \\Holographic Optics}
\author{ Xian-Hui Ge${ }^{1}~$, ~Kwanghyun Jo${}^{2}~$ ~~and  ~~Sang-Jin Sin${}^{2}~$}
\date{}
\maketitle \vspace{-1.0cm}
\begin{center}
~~~
${}^{1}$Department of Physics, Shanghai University, Shanghai 200444, China\\
~{}
${}^{2}$Department of Physics, Hanyang University, Seoul 133-791, Korea\\
~{}
\\
{\small {E-mail: {{gexh@shu.edu.cn}, ~{jokh38@gmail.com}, ~{sjsin@hanyang.ac.kr}}}}
~~~\\
~~~\\
\end{center}
\begin{abstract}
We  consider the AdS$_4$ RN black hole and work out the momentum dependent
hydrodynamic analysis for the vector modes.   We also perform the spectral function calculation of the dual
field theory. As an application, we consider the permittivity and permeability
and find that for low frequency regime, the index of refraction is found to be
negative,  supporting the  claim  made in ref.\cite{amariti} for AdS$_5$.
We also find  that at static limit  the medium has the  zero permeability, a  character of the superconductivity . 
 \end{abstract}

{\small
\begin{flushleft}
\end{flushleft}}
\newpage

\tableofcontents 

\section{Introduction }
The  gauge/gravity duality  \cite{ads/cft,gkp,w} has been
proved to be the most fruitful idea in recent string theory and  has been widely used in several
different systems:  from  the quark-gluon plasma \cite{pss0,kss,bl, ssz,ksz,gssmt,gssmt1,gs} and to condensed matter systems \cite{hartnoll,hartnoll2,sin}. The super conductivity\cite{horowitz}  and transport properties \cite{sachdev} were calculated for the 2+1 dimensional system near the critical  point and dual background of the non-trivial scaling properties were also suggested \cite{Kachru}.  
In a very recent paper \cite{amariti}, a new connection between AdS gravity and optics we suggested: the negative refractive index of some meta-materials can be described by the holographic duality.

The most basic configuration is the asymptotically anti-de Sitter black hole with electric charge, which describe the finite temperature and chemical potential of the dual field theory. For the transport coefficient, one only needs to calculate the Green functions for the zero momentum case and corresponding analysis were done in \cite{hartnoll,hartnoll2}. On the other hand, the permeability request momentum dependent hydrodynamic analysis even for the small frequency limit.

In this paper, we first perform the momentum dependent hydrodynamic
analysis for four-dimensional Reissner-Nordstr$\rm\ddot{o}$m-Anti-de-Sitter black holes. The
decoupled equations of motion and the Green functions have been
worked out explicitly. We then study the spectral functions for the vector modes and
as an application, refractive index for a strongly coupled medium were calculated.
 
In \cite{hartnoll,hartnoll2}, hydrodynamic analysis for the dyonic
black hole for zero momentum case but with magnetic charge was performed. The zero momentum  limit of our result is different from their result by taking the zero magnetic limit. This is because the limit of zero magnetic charge and that of zero momentum is not commuting. It is singular as was argued by the authors of ref. \cite{sachdev}. However our result gives consistent result for the conductivity in the literature \cite{Erdmenger,Hur}. We also calculated the spectral functions, permittivity and permeability and for low frequency regime, the index of refraction is found to be negative supporting the claim  made in ref.\cite{amariti} for AdS$_5$. After finishing this work, we are informed that propagator structure with  general quasi-normal modes has been studied in
a very recent paper  \cite{brattan} and hydrodynamic analysis at zero temperature is described in \cite{jottar}.

\section{Basic setup}
The action for Einstein-Maxwell theory with a negative cosmological
constant $\Lambda=-3/L^2$ is give by \be \label{action}
S=\frac{1}{2\kappa^2_4}\int d^4 x\sqrt{-g}(R-2\Lambda-L^2
F_{\mu\nu}F^{\mu\nu})
+\frac{1}{ \kappa^2_4}\int d^3 x\sqrt{-\gamma}\Theta
+\frac{2}{ \kappa^2_4 L}\int d^3 x\sqrt{-\gamma} ,\ee
where $F_{\mu\nu}$ is the field strength of Maxwell field ${\cal{A}}_\mu$, the second integral contains the Gibbons-Hawking term and the third one contains the counter term for the regularization. Here $\gamma_{\mu\nu}$ is the induced metric on the boundary and $\Theta=\gamma^{\mu\nu}\Theta_{\mu\nu}$ is the trace of the
extrinsic curvature $\Theta_{\mu\nu}=-(\nabla_\mu n_\nu+\nabla_\nu n_\mu)$ with $n$ the unit normal vector to the boundary directed outward.

The 4-dimensional Reissner-Nordstr$\ddot{o}$m-Anti de Sitter black
hole is the solution of Einstein equation and the metric is
\ba
ds^2 &=&\frac{l^2
\alpha^2}{z^2}\left(-f(z)dt^2+dx^2+dy^2\right)+\frac{l^2}{z^2}\frac{dz^2}{f(z)}, \\
A^{(0)}_t&=&-Q\alpha z.
\ea
The function $f(z)$ has the form
\ba
f(z)=1+Q^2z^4-(1+Q^2)z^3.
\ea
The Hawking temperature of the black hole reads
\be
T=\frac{(3-Q^2 )\alpha}{4\pi}.
\ee
We consider the linearized perturbations $h_{tx}$, $h_{xy}$ and $A_x$ of the
Einstein equation and the Maxwell equation, $g_{\mu\nu}=g^{(0)}_{\mu\nu}+h_{\mu\nu}$, ${\cal{A}}_\mu = A^{(0)}_\mu + A_\mu$. From the Maxwell equation we find
\ba
f(fA'_x)'+\omega^2A_x-Qf{h^{x}_t}'
-k^2_yfA_x=0. \label{Ax}
\ea
The linearized equations from the Einstein equation yield
\ba \label{Einsteinvcteom1}
&&- {{h^{x}_t}''} +\frac{2}{ z }  {h^{x}_t}' +4Q z^2A'_x
+\frac{k^2_y}{f}h^{x}_t+\frac{\omega k_y}{ f}h^{x}_y=0,\label{hx}\\
&&  \omega  {h^{x}_t}'  - 4\omega  Qz^2 A_x+ { k_yf}{h^{x}_y}'=0,\label{ht}\\
&&{h^x_y}''+\frac{(z^{-2}f)'}{z^{-2}f}{h^x_y}'+\frac{\omega}{f^2}\left( k_y
h^x_t+\omega h^x_y\right)=0.
\ea
In these equations we have transformed
the variables $A_x\rightarrow A_x \alpha $, $\omega\rightarrow \omega \alpha $  and
$k\rightarrow k\alpha$, so that all the coordinate, momentum as well as field variables are dimensionless. Therefore that $h^{x}_t  $ and ${h^x_y} $ were defined to be dimensionless from the beginning.  At the end of the calculation, we can restore the original variables by reversing the scaling.

In order to have a decoupled equation, we define a new variable
\ba
\Phi_{\pm}=\frac{{h^{x}_t}'}{ z^2}-4QA_x+\frac{C_{\pm}}{ z}A_x .
\ea
Then equations (\ref{Ax}), (\ref{hx}) and (\ref{ht}) reduce to two decoupled equations for $\Phi_\pm$
\ba\label{mas}
\Phi''_{\pm}+\frac{(z^{2}f)'}{z^{2}f}\Phi'_{\pm}+\frac{\omega^2-k^2f}{f^2}\Phi_{\pm}
-\frac{C_{\pm}Q z}{f}\Phi_{\pm}=0, \ea
where $C_{\pm}=\frac{3+3Q^2\pm\sqrt{9(Q^2+1)^2 +16k^2Q^2}}{2Q}$.

\section{Hydrodynamic analysis}
We now seek a transformation
\be \Phi(z) = g(z) \Xi(z) \ee
such that the coefficient of $\Xi(z)$  goes to zero when we take the hydrodynamic limit, $\omega,k\to 0$.
Here $\Phi$ represent one of $\Phi_\pm$.
 \be
\left(\frac {\omega^2}{f(z)^2}-\frac{k^2 + {C_{\pm}} Q z}{f(z)} + \frac {2 g'(z)} {z g(z)} +\frac{f'(z) g'(z)}{g(z)f(z)} +\frac{g''(z)}{ g(z)} \right) \Xi(z) + \left (\frac {2} {z} + \frac {f'(z)} {f(z)} + 2 \frac{g'(z)}{g(z)} \right)\Xi'(z)+\Xi''(z) = 0 \label{eqpsi}
\ee
We need to find $g$ such that $\Xi$'s coefficient in eq.(\ref{eqpsi}) is of order $k^2$ or higher in hydrodynamic limit. The constant $C_{\pm}$ can be expanded in a series of $k^2$
\ba
&& C_+= \frac {3\left (1 + Q^2 \right)}{Q} + \frac {4 Q k^2}{3\left (1 + Q^2 \right)}
+\mathcal{O}(k^4) \\
&& C_{-}=-\frac{4 Q k^2}{3 \left(1+Q^2\right)}+\mathcal{O}(k^4).
\ea
\subsection{For $ \Phi _+$ }
We request that $g(z)$ satisfies following equation
\be \label{gz}
-\frac {3\left (1 + Q^2 \right) z } {f(z)} g(z)+
\left(\frac {2} {z} + \frac {f'(z)} {f(z)} \right) g'(z) + g''(z) = 0.
\ee
The  solution for (\ref{gz}) regular at $z=1$ is given by
\be
g(z)=\frac{1}{z}-  \frac{4Q^2}{3(1+Q^2)} ,
\ee
where we used the linearity of equation of motion to normalize the solution.

By imposing the infalling boundary condition and taking out the near horizon limit,
\be
\Xi(z) =(1 - z)^{\nu} F(z),
\ee
with $\nu=-i\omega/4\pi T$.
\footnote{Here both $\omega$ and $T$ are scaled by $\alpha$ to be dimensionless:  $  T  =(3 - Q^2)/4\pi$ and $\nu = -i  {\omega}/{(3 - Q^2)}$.}
\ba
F(z) \left(\frac{( \nu-1 ) \nu }{(1-z)^2}+\frac{\omega^2-k^2 \left(1+\frac{4 Q^2 z}{3 \left(1+Q^2\right)}\right) f(z)}{f(z)^2}-\frac{\nu}{1-z} \left(\frac{f'(z)}{f(z)}+ \frac{2}{z}+\frac{2g'(z)}{g(z)} \right) \right)\\
+F'(z) \left(\frac{-2 \nu }{1-z}+ \frac{f'(z)}{f(z)}+ \frac{2}{z}+\frac{2g'(z)}{g(z)} \right)+F''(z)=0
\ea
In the long-wavelength, low-frequency limit, we can expand $F(z)$ in
a double series with respect to $\omega$ and $k$
\ba
F(z) &=& C_0 + \omega {F_1}(z) + k^ 2 F_2(z) +\mathcal {O}(\omega^2,k^2) \\ \no
0 & = &  {F_1}''(z) + {F_1}'(z)\left (\frac {f'(z)} {f(z)} + \frac{2}{z}+\frac{2g'(z)}{g(z)} \right) \no
&& +\frac{1}{3-Q^2}\left(\frac {i C_0} {(-1 + z)^2} + \frac {2 i C_0} {z -
z^2} + \frac {i C_0 f'(z)} {f(z) (1 - z)} + \frac {2 i C_0 g'(z)} {g(z) (1 - z)} \right)  \\
{F_1} &=& i C_0\int_ 1^z \left (\frac {1} {(3-Q^2)( x-1)} +
\frac {g(1)^2}{f(x) x^2g(x)^2} \right) \, dx := i C_0 H(z) \label{H(z)} \\ \no
0 &=& {F_2 }''(z)-+ {F_2 }'(z)\left (\frac {f'(z)} {f(z)} +\frac{2}{z}+\frac{2g'(z)}{g(z)} \right)\frac {C_0} {f(z)}\left (1 + \frac {4 Q^2 z} {3\left (1 + Q^2 \right)} \right) \\
{F_2 } &=& \frac {C_0} {3\left (1 + Q^2 \right)}\int_ 1^z\frac {\int_ 1^y x^2\left (3 + Q^2 (3 + 4 x) \right) g(x)^2\,dx} {y^2 f(y) g(y)^2}\, dy := C_0 J(z)  \label{J(z)}
\ea
Later we will need following results for $H(z)$ and $J(z)$:
\ba
H(z)&=&H(0)+H_1 z+ \cdots , \quad J(z)=J(0)+J_1 z+ \cdots \no
\mbox{with} && H_1 = \frac{18-45 Q^2-Q^6}{9 (3-Q^2)\left(1+Q^2\right)^2},\\ \label{H1}
J_1&=& -\frac{27+63 Q^2+29 Q^4+9 Q^6}{27 \left(1+Q^2\right)^3}. \label{J1}
\ea

\subsection{For $\Phi_-$ }
Since $C_- \sim O(k^2) $,  we do not need transformation factor and we set  $g(z) =
1$, unlike $\Phi_+$ case. By doing as before, we get
\begin{eqnarray}
\left( \frac{\nu (-2+z+z \nu)} {z(z-1)^2} +\frac{\nu f'(z) } {(z-1) f(z)}
+\frac{ \left(w^2-k^2f(z)\left(1-\frac{4Q^2z}{3(1+Q^2)} \right)\right)}{f(z)^2} \right)F(z) \nonumber \\
+\left(\frac{2}{z}   +\frac{f'(z)} { f(z)} +\frac{2\nu}{z-1}\right) F'(z) +F''(z)=0
\end{eqnarray}
Expanding  $F(z) = \tilde{C}_0 + \omega {F_1}(z) + k^2 {F_2}(z) +\mathcal{O}(\omega^2,k^2)$, we have
\ba
\frac{i \tilde{C}_0}{(3-Q^2)(1-z)}\left(\frac{1}{1-z}+\frac{2}{z}+ \frac{f'(z)}{f(z)} \right)  +\left (\frac{  2}{z}+ \frac{f'(z)}{f(z)} \right) {F_1}'(z) + {F_1}''(z) &=&0,\\
-\frac{\tilde{C}_0}{f(z)}\left(1-\frac{4 Q^2 z}{3 \left(1+Q^2\right) }\right)+F_2'(z) \left(\frac{f'(z)}{f(z)} +\frac{2}{z}\right) +F_2''(z)&=&0
\ea
and their solutions are
\ba
{F_1}&=&i \tilde{C}_0\int_1^z \left(\frac{1}{(3-Q^2)( x-1)}+\frac{ 1}{f(x) x^2}\right) \, dx \\
{F_2}&=&-\frac{\tilde{C}_0 }{3 \left(1+Q^2\right)}\left(1-\frac{1}{z}\right)
\ea
respectively.

Let us fix the constants $C_0$ and $\tilde{C}_0$ by imposing boundary conditions as
\be
{\lim_{z\rightarrow 0}}h^x_t(z)=\hat{h}^x_t, ~~~\lim_{z\rightarrow
0} h^y_x(z)=\hat{h}^x_t,~~~\lim_{z\rightarrow 0}A_{x}=\hat{A}_x
 \ee
It would be simple by taking a derivative of $\Phi_{\pm}(z)$ and
using equation (\ref{hx}). This gives the relation
\be
z^2 \Phi'_{\pm}-C_{\pm}z A'_x=\frac{1}{f(z)}
\bigg(k^2 h^x_t+\omega k h^y_x\bigg)-C_{\pm}A_x.
\ee
In order to determine $C_0$ and $\tilde{C}_0$, we need to examine the above equation at the boundary
\be
\lim_{z\rightarrow 0} (z^2 \Phi'_{\pm}-C_{\pm}zA'_x)=\bigg(k^2
\hat{h}^x_t+\omega k \hat{h}^x_t\bigg)-C_{\pm}\hat{A}_x
\label{BC} :=L_\pm
\ee
We have a few remarks on eq. (\ref{BC}).
\begin{itemize}
\item
Only $1\over z$ singularity of $\Phi$  near boundary can contribute to the left hand side.
\item The boundary value of $z A_x$ is 0.
\item
The equation (\ref{BC}) is correct even for the finite $k$ and $\omega$.
\end{itemize}
Let
\be
\Phi_\pm =\frac{\hat{\Phi}_\pm}{z} + \hat{\Pi}_\pm + T_\pm \log z \cdots \label{RST}
\ee
 be the expansion near the boundary. Then
\ba
\hat{\Phi}_+&=&C_0\left(1+i\omega H(0)+k^2 J(0) \right) =-L_+,\quad \\
\hat{\Phi}_-&=&{\tilde C}_0\left(- i\omega  +\frac{k^2}{3(1+Q^2)}\right) =-L_-.
\ea
and
\ba
\hat{\Pi}_+ &=& -L_+\left(\frac{i\omega}{3-Q^2} -\frac{4}{3}\frac{Q^2}{1+Q^2} +i\omega H_1+k^2 J_1 \right)  + {\cal O}(k\omega,\omega^2) \\
\hat{\Pi}_- &=& {\tilde C}_0 + {\cal O}(k\omega,\omega^2) \\
T_\pm&=&0
\ea
where $H_1,J_1$ are given in eq. (\ref{H1}) and eq. (\ref{J1}) respectively.

The constants $C_0$ and $\tilde{C}_0$ can be read off from the $\hat{\Phi}_\pm$ and $L_\pm$ parts,
respectively
\ba
C_0 &=& \frac{-\bigg(k^2 \hat{h}^x_t+\omega k
\hat{h}^x_t\bigg) +\left(\frac {3\left (1 + Q^2 \right)} {Q} + \frac
{4 Q k^2} {3\left (1 + Q^2 \right)}\right)\hat{A}_x}{1+i\omega H(0)+k^2 J(0) } \\
\tilde{C}_0&=&\frac{\bigg(k^2 \hat{h}^x_t+\omega k
\hat{h}^x_t\bigg)+\frac{4 Q k^2}{3 \left(1+Q^2\right)}\hat{A}_x}{ i\omega - \frac{k^2}{3(1+Q^2)} } ,
\ea
where $H(z)$ and $J(z)$ were defined at eq. (\ref{H(z)}) and eq. (\ref{J(z)}) respectively.

Let us turn to the $A'_x(z)$ and ${h^x_t}'$ near the boundary.
From the definition of the master fields, we can solve
${h^x_t}'$ and $A_x(z)$ in terms of the $\Phi_\pm$:
\be
{h^x_t}'=z^2\Phi_{-} +z^2(4Qz -C_{-})\frac{\Phi_{+}-\Phi_{-}}{C_{+}-C_{-}}, \quad \quad A_x=z\frac{\Phi_{+}-\Phi_{-}}{C_{+}-C_{-}}.
\ee
For example $ A'(\epsilon)$ can be calculated from the observation:
\be
A'(\epsilon)=\frac{\hat{\Pi}_+ -\hat{\Pi}_-}{C_+ - C_-}+ {\cal O}(k\omega,\omega^2,\epsilon)
\ee
where $\hat{\Pi}_\pm$ are defined in eq. (\ref{RST}). Now, we can evaluate the solutions for the fields near the boundary $z=\epsilon$:
\ba
&& {h^x_t}' = -\frac{\epsilon}{\alpha^2} \bigg(k^2 \hat{h}^x_t+\omega k
\hat{h}^x_t\bigg)+ \frac{\epsilon^2}{\alpha}\cdot\frac{ k^2
\hat{h}^x_t+ \omega k \hat{h}^x_t +4i\omega Q \hat{A}_x}{i\omega -D k^2} + \mathcal {O}(\epsilon^3),\\
&& {h^y_x}'=\frac{\epsilon}{\alpha^2}\bigg(k\omega \hat{h}^x_t+\omega^2
\hat{h}^x_t\bigg) -\frac{\epsilon^2}{\alpha}\cdot\frac{ k\omega\hat{h}^x_t+ \omega^2\hat{h}^x_t + 4QDk\omega  \hat{A}_x}{i\omega -D k^2} + \mathcal {O}(\epsilon^3),\\
&& A'(\epsilon )=  {-Q}{D\alpha}\cdot \frac{\bigg(k^2 \hat{h}^x_t+\omega k \hat{h}^x_t+4i\omega {Q} \hat{A}_x\bigg)}{i\omega -D k^2}  + {4Q D^2 \alpha k^2} \left(h_t^x\right)^0 \\
&& ~~~~~~~~~~~~~+ \hat{A}_x\left( i\omega\frac{(3-Q^2)^2}{9\alpha(1+Q^2)^2}
-\frac{9+21 Q^2-Q^4+3 Q^6}{ 9 \left(1+Q^2\right)^3\alpha^2} k^2 \right) + \mathcal {O}(\epsilon^3).
\ea
Notice that $\alpha$ factor were restored and $D$ is the diffusion constant
\be
D= \frac{1}{3\alpha(1+Q^2)}.
\ee

For the Green functions, we need on-shell action that gives finite and quadratic function of the boundary values. Using the fact
\be
\delta S=\int\partial_{\mu}\bigg(\frac{\partial{L}}{\partial\partial_{\mu}\phi^i}\delta\phi^i\bigg)dz
+\int E.O.M. \delta\phi^i dz,
\ee
and deleting all the contact terms, it is given by
\be \label{bdryact}
S= \frac{l^2 \alpha}{\kappa^2_4}\int  \frac{d^3k}{(2\pi)^3}
\left[\frac{\alpha^2}{4z^2}\bigg(h^x_t{h^x_t}' -f(z)h^y_x{h^y_x}'\bigg)-f(z) AA'  +Q\alpha A h_t^x\right]_{z\to 0}^1+\cdots
\ee
The factor $1\over4$ comes from $-3/4+1$ where $-3/4$ is from the original action while $+1$ is from the Gibbons Hawking term.

Following ref. \cite{sonstarinets1},  we  discard the contribution of the
horizon as a prescription for computing thermal Green's functions in Minkowski space.
From the definition, $G_{i;j}=-\delta^2S/\delta{\phi_i}\delta{\phi_j}$, one finds the
correlators in the hydrodynamic approximation
\ba
G_{xt;xt}&=&\frac{l^2\alpha^2}{2\kappa^2_4}
\cdot \frac{k^2}{i\omega-Dk^2} ,\quad
G_{xy;xy} = \frac{l^2\alpha^2}{2\kappa^2_4}  \cdot\frac{\omega
^2}{i\omega-Dk^2} ,\\
G_{xt;xy}&=&\frac{l^2\alpha^2}{2\kappa^2_4} \cdot \frac{\omega
k}{i\omega-Dk^2} ,\\
G_{xt;x}&=& \frac{l^2\alpha^2}{ \kappa^2_4}\left(\frac{ 2i\omega Q}{i\omega-Dk^2} -4Q D^2k^2\right),\\
G_{xy;x}&=&\frac{l^2\alpha^2}{ \kappa^2_4} \cdot\frac{2Q D k\omega
}{i\omega-Dk^2} ,\\
G_{x;x}&=&
\frac{2l^2\alpha }{\kappa^2_4}   \left(\frac{ 4i \omega\alpha Q^2 D }{i\omega-D
k^2} -i\omega \frac{(3-Q^2)^2}{9\alpha(1+Q^2)^2} + k^2\frac{(9+21 Q^2-Q^4+3 Q^6)}{ 9 \alpha^2\left(1+Q^2\right)^3}  \right).\ea

The transport coefficients can be read off from the green function. The DC conductivity $\sigma$
is given by
\be \label{condKubof}
\sigma =-\lim_{\omega\to 0}\frac1{\omega} \mbox{Im} G_{xx}(\omega,k=0)
=\frac{l^2}{\kappa^2_4} \frac{(3-Q^2)^2}{9(1+Q^2)^2} .
\ee
Notice that this is dimensionless and vanishes as square of the temperature
( $\sim T^2$)  near the zero temperature while it is constant in high temperature.

\section{Spectral function}
To calculate spectral function of current-current operator, we need to know the boundary value of the fields and theirs conjugate momentum. For the numerical computation, it is better to introduce new master variable $\Psi_\pm$ in this section which is different from $\Phi_\pm$ by a factor 1/z,
\be
\Psi_\pm = z \Phi_\pm = \frac{1}{z} {h^x_t}' +\left(C_\pm - 4Q z \right) A_x. \label{4.1}
\ee
We remark that the new master field $\Psi_\pm$ has nothing to do with $\Psi$ in eq.(3.1).
The equation of motion   in terms of the new master variables are
\be
\Psi_\pm'' + \frac{f'}{f} \Psi_\pm' +\frac{1}{f}\left(\frac{\omega^2-k^2 f}{f}-C_\pm Q z -\frac{f'}{z} \right)\Psi_\pm=0, \label{4.2}
\ee
where $C_\pm$ is
\ba
C_{\pm}=\frac{3+3Q^2\pm\sqrt{9(Q^2+1)^2 +16k^2Q^2}}{2Q}=\frac{3(1+Q^2)}{2Q}\left(1 \pm \sqrt{1+\left(\frac{4Qk}{3(1+Q^2)}\right)^2}\right)
\ea
From the  equations for master field we can get the spectral function of master fields. We however need the spectral function of original variables not the master fields itself. So let us first find the series solution of $h_t^x, A_x$ which defines the conjugate momentums $\pi_h,\pi_a$
\ba
h^x_t &=& \hat{h}_t^x -\frac{k}{2}\hat{Z}_1 z^2 +\frac{1}{3}\hat{\pi}_h z^3 + \cdots \no
A_x &=& \hat{A}_x + \hat{\pi}_a z + \cdots ,
\ea
where $\hat{Z}_1 = k \hat{h}_t^x + \omega \hat{h}_y^x$. The coefficient $\alpha_i, \beta_i$ is computed from eq. (\ref{Einsteinvcteom1}) : The master variables have series solution near the boundary,
\be \label{masterseriessol}
\Psi_\pm = \hat{\Psi}_\pm+ \hat{\Pi}_\pm z +\cdots.
\ee
The eq. (\ref{4.1}) can be written as a matrix form \cite{Jo:2010sg}
\be
\left(\begin{array}{c}
\Psi_+\\
\Psi_-
\end{array}\right) = \mbox{R} ~
\left(\begin{array}{c}
{h^x_t}'\\
A_x
\end{array}\right), \;\; \mbox{with} \;\;
\mbox{R}= \left(
\begin{array}{cc}
1/z & C_+ - 4Qz\\
1/z & C_- - 4Qz
\end{array} \right).
\ee
Similarly,
if  we introduce the transformation matrix R$_0$ ,
\be
\mbox{R}_0= \left(
\begin{array}{cc}
1 & C_+ \\
1 & C_-
\end{array} \right),
\ee
the boundary values and of master fields and those of original fields are related by $R_0$.
\be \label{pimTopio}
\left(
\begin{array}{c}
\hat{\Psi}_+ \\ \hat{\Psi}_-
\end{array} \right) = \mbox{R}_0\left(
\begin{array}{c}
-k\hat{Z}_1 \\ \hat{A}_x
\end{array} \right),\quad
\left(
\begin{array}{c}
\hat{\Pi}_+ \\ \hat{\Pi}_-
\end{array} \right)
=\left(
\begin{array}{c}
\hat{\Psi}_+ \cal{G}_+ \\ \hat{\Psi}_- \cal{G}_-
\end{array} \right)
=\mbox{R}_0
\left(
\begin{array}{c}
\hat{\pi}_h \\ \hat{\pi}_a
\end{array} \right)
\ee
we use the linear response relation $\hat{\Pi}_\pm = \hat{\Psi}_\pm \cal{G}_\pm$ in eq. (\ref{pimTopio}). Then the conjugate momentum $\hat{\pi}_h, \hat{\pi}_a$ are written as
\ba
&& \left(
\begin{array}{c}
\hat{\pi}_h \\ \hat{\pi}_a
\end{array} \right)
=\Lambda
\left(
\begin{array}{c}
\cal{G}_+ \\ \cal{G}_-
\end{array} \right), \quad \mbox{where} \quad \Lambda = \mbox{R}_0^{-1} \mbox{Diag}(\Psi_+,\Psi_-) \no
\Lambda &=& \frac{1}{C_+ - C_-}\left(\begin{array}{cc}
C_-(k \hat{Z}_1 - C_+ \hat{A}_x) & C_+(-k \hat{Z}_1 + C_- \hat{A}_x) \\
-k \hat{Z}_1 + C_+ \hat{A}_x & k \hat{Z}_1 - C_- \hat{A}_x
\end{array} \right)
\ea
From the boundary action eq.(\ref{bdryact}), we compute two point function in terms of boundary values ($\hat{h}^x_t, \hat{A}_x$) and conjugate momentum of master field $\hat{\Pi}_\pm$:
\be
G_{xtxt}= \frac{\delta^2 S_{bd}}{\delta \hat{h}^x_t \delta \hat{h}^x_t} ,\quad G_{xyxy}= \frac{\delta^2 S_{bd}}{\delta \hat{h}^x_y \delta \hat{h}^x_y} ,\quad G_{xx}=  \frac{\delta^2 S_{bd}}{\delta \hat{A}_x \delta \hat{A}_x}
\ee
The two point function for $h^x_t$ and $h^x_z$ is related by Ward identity,
therefore the correlation functions for each components are
\ba \label{GREENFNC1}
\frac{2\kappa_4^2}{l^2 \alpha^3}G_{xt,xt}&=&\mathcal{G}_{xt,xt}= k^2 \frac{C_- {\cal{G}}_+-C_+{\cal{G}}_-}{C_+-C_-} ,\quad
\frac{2\kappa_4^2}{l^2 \alpha^3} G_{xy,xy}=\mathcal{G}_{xy,xy} =\frac{\omega^2}{k^2} \mathcal{G}_{xt,xt}\no
\frac{\kappa_4^2}{l^2 \alpha^3}G_{xt,x}&=& \mathcal{G}_{x,xt} = k^2 \frac{{\cal{G}}_+-{\cal{G}}_-}{C_+-C_-} \no
\frac{\kappa_4^2}{l^2 \alpha}G_{xx}&=&\mathcal{G}_{x,x} = \frac{C_+ {\cal{G}}_+ - C_- {\cal{G}}_-}{C_+-C_-}.
\ea
\begin{figure} \
\begin{center}
    \includegraphics[angle=0, width=0.43 \textwidth]{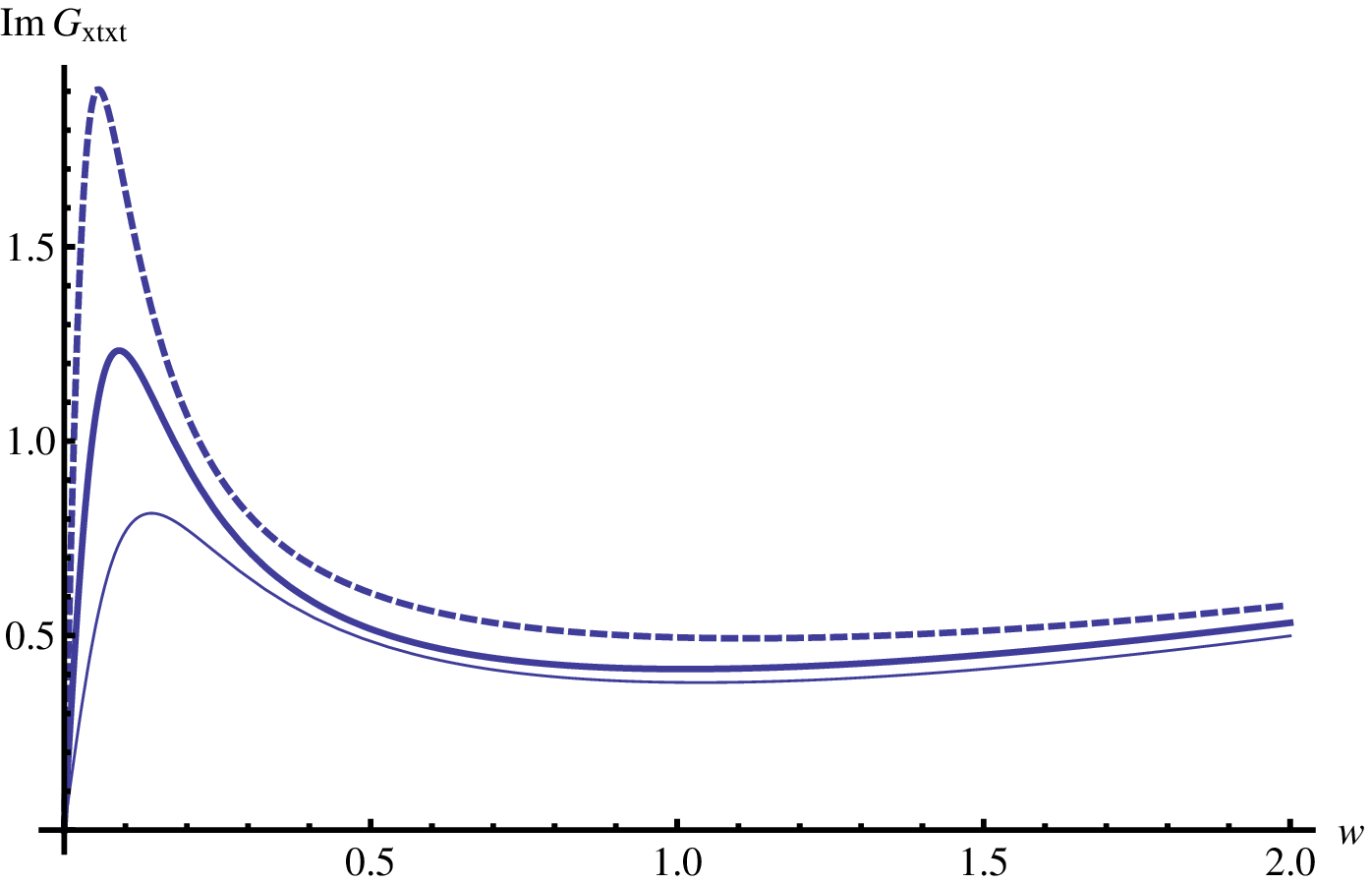}
    \includegraphics[angle=0, width=0.43 \textwidth]{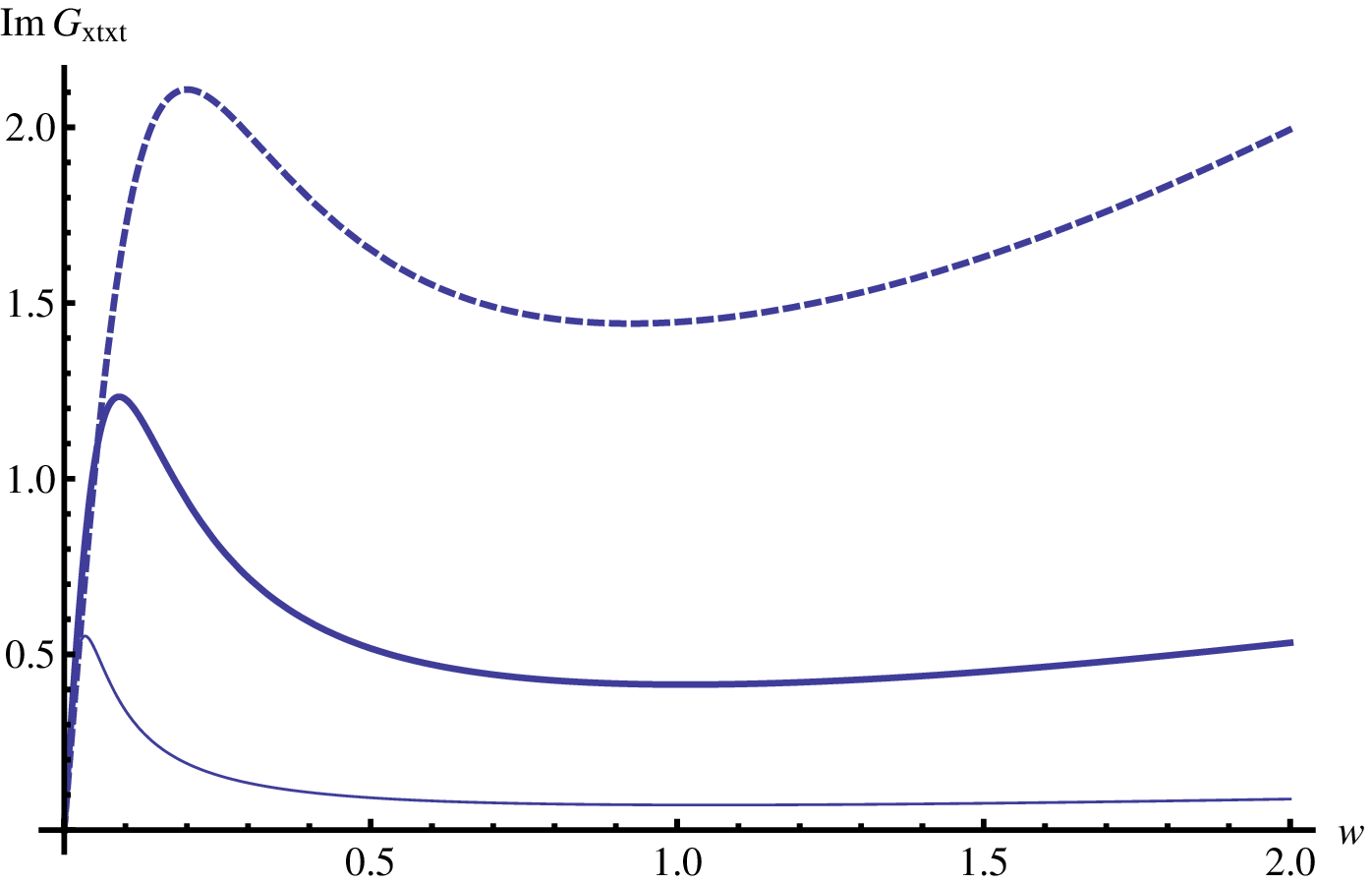}
  \caption{Spectral function of xt,xt component, Im ${\cal{G}}_{xtxt}$. Left : with fixed k=0.5 varying Q=0.5(thin), 1(thick), 1.5(dashed). Right : with fixed Q=1, varying k=0.2(thin), 0.5(thick), 1(dashed)}
\end{center}
\end{figure}

\begin{figure} \
\begin{center}
    \includegraphics[angle=0, width=0.43 \textwidth]{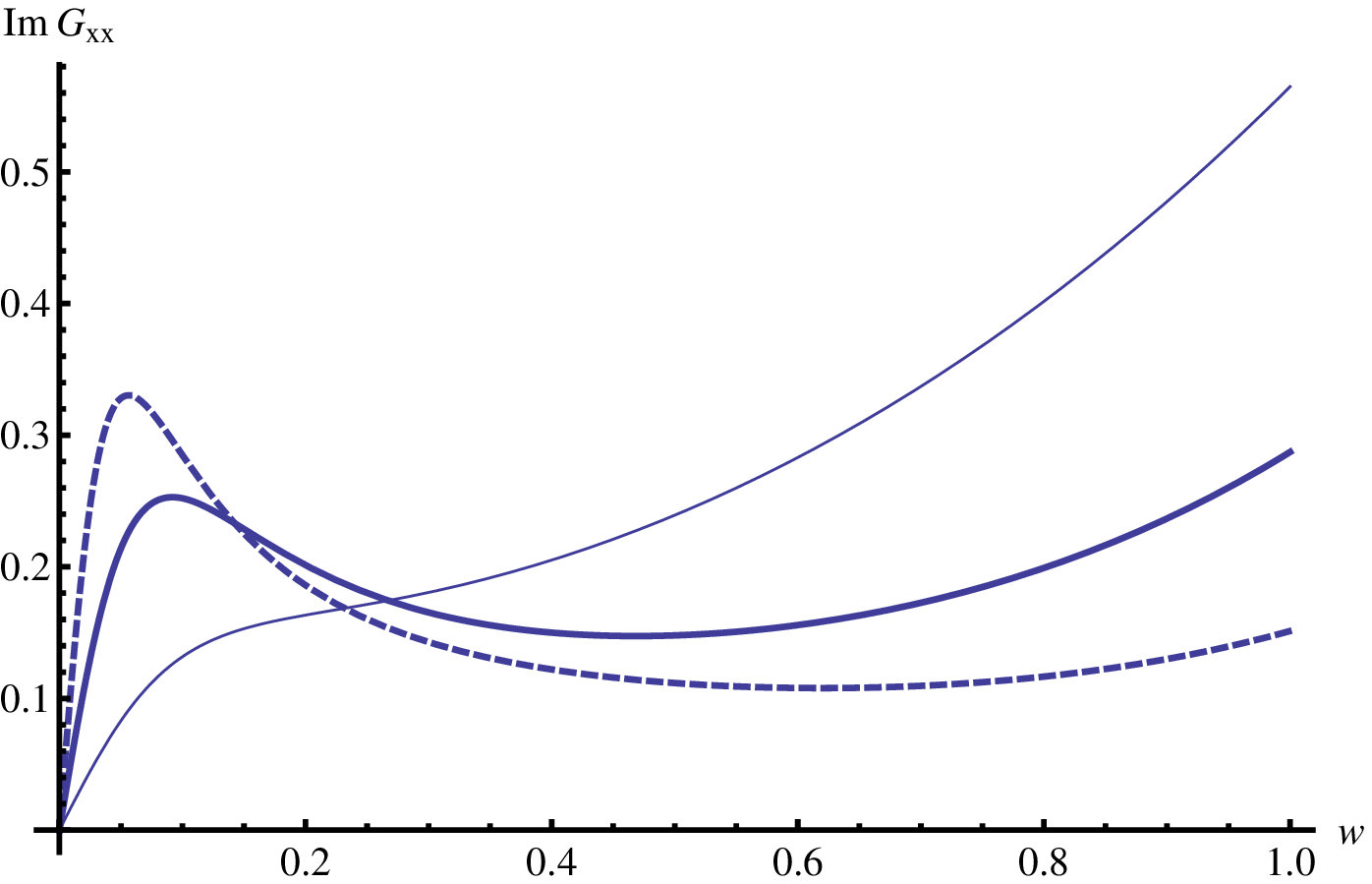}
    \includegraphics[angle=0, width=0.43 \textwidth]{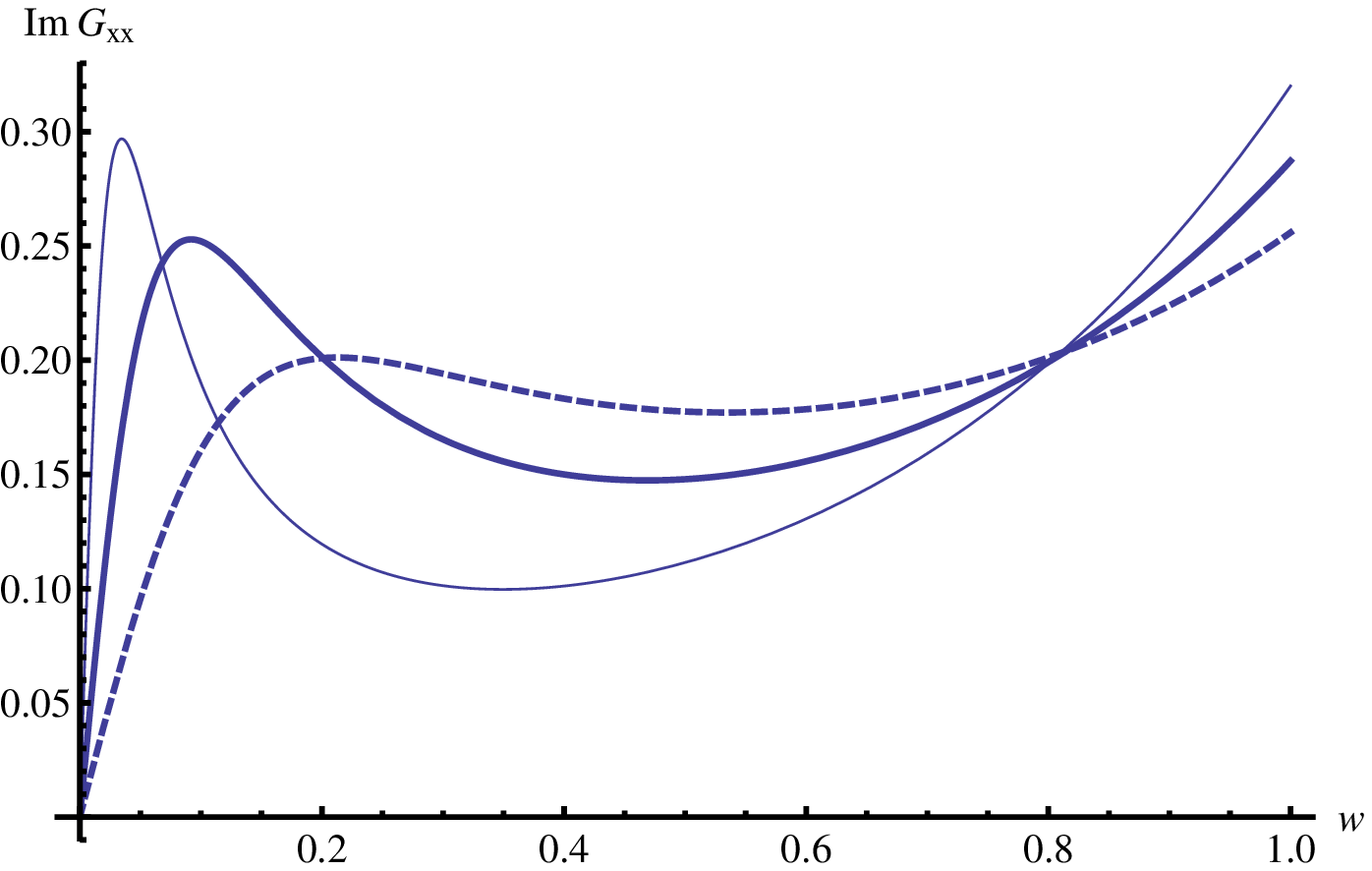}
  \caption{Spectral function of xx component, $\mbox{Im} {\cal{G}}_{xx}$. Left : with fixed k=0.5 varying Q=0.5(thin), 1(thick), 1.5(dashed). Right : with fixed Q=1, varying k=0.2(thin), 0.5(thick), 1(dashed)}
\end{center}
\end{figure}

\subsection{holographic recipe for Green function}
In general, the second order differential equation has two independent solution and these two independent solution have two different integration constant. We represent the near the boundary solution by $\Phi_i$ and near horizon solution by $\phi_i$. Then,
\ba
\Phi_1 &=& u^{\Delta_-}(1 + \cdots), \quad \phi_1 = (1-u)^{-\nu} (1+ \cdots) \no
\Phi_2 &=& u^{\Delta_+} (1+ \cdots), \quad \phi_2 = (1-u)^{\nu} (1+ \cdots)
\ea
where u = 0 is the UV boundary or AdS boundary and u=1 is IR boundary or black hole horizon when we consider finite temperature system. Near the boundary, u=0, $\Phi_1$ and $\Phi_2$ are the local solutions and $\Delta_\pm$ is the solution of indicial equation near the boundary, $\Delta_+ > \Delta_-$ where $\Delta_+$ is the conformal dimension of an operator and $\Delta_-$ is the dimension of the dual source field. Near the horizon, u=1, $\phi_1, \phi_2$ are the local solution and they are identified with infalling and outgoing solution respectively.
\ba
& & \mathcal{A}(\wn,\qn)\Phi_1 +\mathcal{B}(\wn,\qn)\Phi_2 = \mathcal{C}(\wn,\qn)\phi_1 +\mathcal{D}(\wn,\qn)\phi_2 \no
E_\alpha &=& \phi_1(u) = \mathcal{A}(\wn,\qn) \Phi_1(u) +\mathcal{B}(\wn,\qn) \Phi_2(u).
\ea
where $\mathcal{A}(\wn,\qn)$ and $\mathcal{B}(\wn,\qn)$ are two integration constants and will be identified with the boundary value of the bulk field and its conjugate variable, respectively. To get the Green function it is first to evaluate the on-shell action,
\ba \label{onshellaction}
S_{\mbox{on-shell}}   
& =& \lim_{\epsilon \rightarrow 0} \int d^d x \sqrt{g} g^{\alpha \beta}{E_\alpha} {E_\beta}'|_{u=\epsilon}
\no
& = & \lim_{\epsilon \rightarrow 0} \int d^{d} x \left[u^{-(\Delta_+ + \Delta_- -1)} \calA^2  \frac{\calA \Delta_- u^{\Delta_- -1}+\Delta_+ \calB u^{\Delta_+ -1}}{\calA u^{\Delta_-}+\calB u^{\Delta_+}} \right]_{u = \epsilon} \no
& = & \lim_{\epsilon \rightarrow 0} \int d^{d} x \calA^2 \left[\Delta_- u^{\Delta_- -\Delta_+} + \Delta_+ \frac{\calB}{\calA}\right]_{u = \epsilon}.
\ea
The first term in the last line is obviously divergent and it should be renormalized by introducing counter term. The second term is Green function which is obtained by differentiating on-shell action with boundary value.

More specifically, the local Frobenius solutions of our master variable near the boundary is
\ba
\Psi_\pm(z,\omega,k) &=& \hat{\Psi}_\pm(\omega,k) \Psi_\pm^A(z,\omega,k) + \hat{\Pi}_\pm(\omega,k) \Psi_\pm^B(z,\omega,k) \no
\Psi_\pm^A(z,\omega,k) &=& 1 + \alpha_1 z + \alpha_2 z^2 + \cdots , \quad \Psi_\pm^B(z,\omega,k) = z(1+\beta_1 z + \beta_2 z^2 + \cdots)
\ea
What we want to calculate is Green function of each field and it is identified with conjugate momentum of boundary field \cite{pss0}. In eq. (\ref{onshellaction}), $\Delta_-$=0 for our master variable, so that kind of divergence is absent. Then
\be \label{GFMSV1}
\frac{\hat{\Pi}_\pm(\omega,k)}{\hat{\Psi}_\pm(\omega,k)}  = \frac{\deriv_z \phi_1(w,k,z)}{\phi_1(w,k,z)}\Bigg|_{z=0}
\ee
Eq. (\ref{GREENFNC1}) and eq. (\ref{GFMSV1}) give us Green functions of original fields. Only remaining is to compute $\Pi_\pm$ numerically.

\subsection{small k expansion}
\begin{figure} \
\begin{center}
    \includegraphics[angle=0, width=0.43 \textwidth]{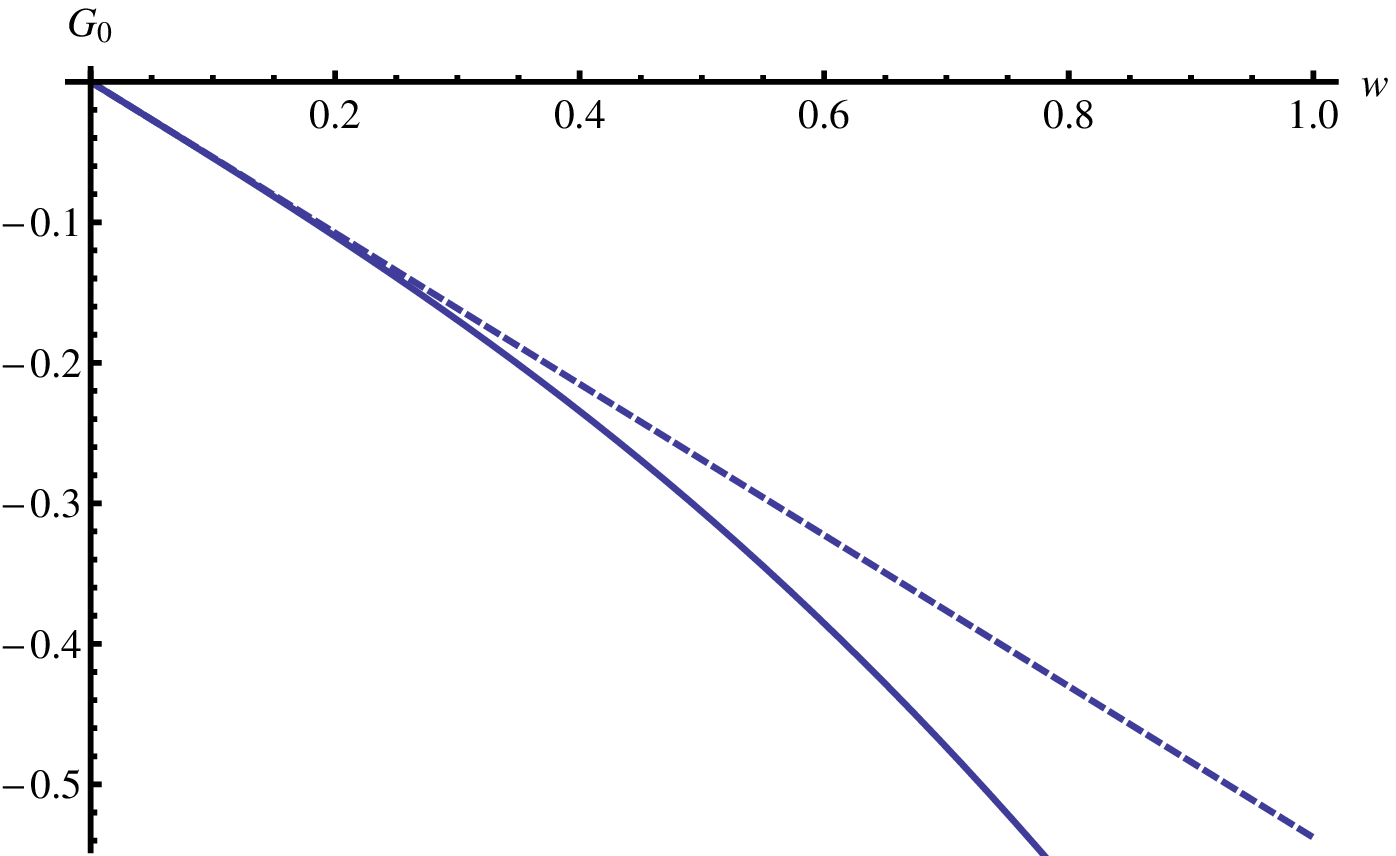}
    \includegraphics[angle=0, width=0.43 \textwidth]{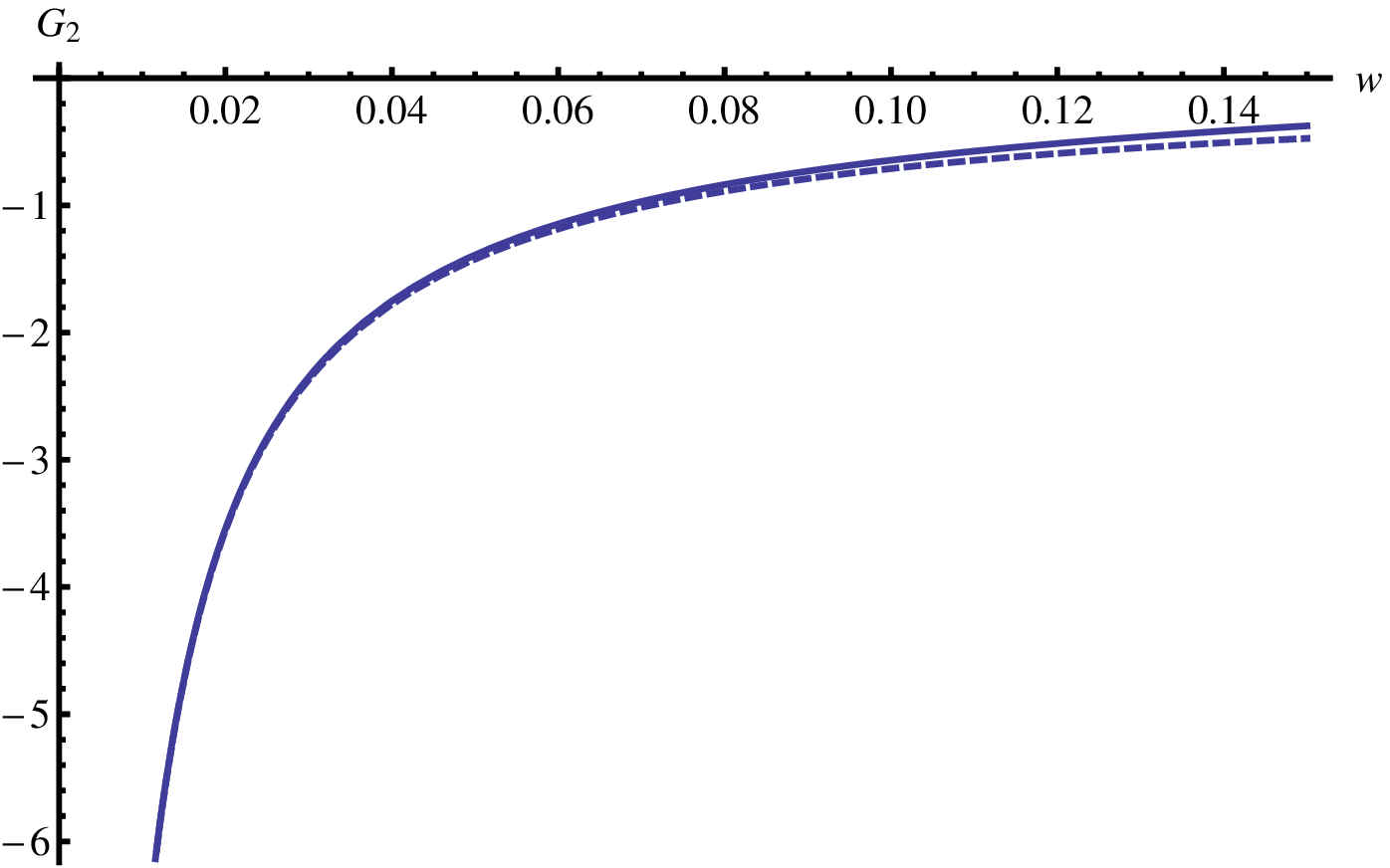}
  \caption{Comparison between numerically (solid) and analytically computed in hydrodynamic limit (dashed) spectral function with Q=0.5. Left is for $G^{0}$(w) and right for $G^{0}$(w). These two functions are well matched in small w region but deviated with each other. }
\end{center}
\end{figure}
In this subsection, we will show how to compute $G_{xx}^{(2)}$ with our previous master field. From eq. (\ref{GREENFNC1}), the expression for current-current Green function is
\be
\mathcal{G}_{xx} = \frac{C_+ {\cal{G}}_+ - C_- {\cal{G}}_-}{C_+-C_-}
\ee
and $G_{xx}$ is expanded near k=0 as $G_{xx} = G^{(0)} +k^2 G^{(2)} + \cdots $, so
\be
\mathcal{G} = \mathcal{G}^{(0)} + k^2 \mathcal{G}^{(2)} + \cdots
\ee
where ${\cal{G}}^{(0)}$ = ${\cal{G}_+} (w,k=0)$ because $C_- (0) = 0$ and ${\cal{G}}^{(2)}$ is
\be
G^{(2)} = \frac{C_-'(0)}{C_+(0)}\Bigg({\cal{G}}_+(0)-{\cal{G}}_-(0)\Bigg)+{\cal{G}}_+'(0).
\ee
To find second order solution for ${\cal{G}}_\pm$, we also need to expand ${\cal{G}}_\pm$ itself as
\be
{\cal{G}}_\pm (w,k) = {\cal{G}}_\pm (w,k=0) + k^2 \frac{\deriv {\cal{G}}_\pm (w,k)}{\deriv k^2}\Bigg|_{k=0} + \cdots = {\cal{G}}_\pm^{(0)} (w) + k^2 ~ {\cal{G}}_\pm^{(2)}(w)
\ee
First we find the infalling solution of the master variable which we call $\phi^{(0)}$ and expand it in momentum, k:
\be
\Psi_\pm = \Psi_\pm^{(0)} + k^2 \Psi_\pm^{(2)} + \cdots = \phi^{(0)}_1 + k^2 \phi^{(2)}_1 + \cdots.
\ee
we then plug it into the eq. \ref{4.1} and expand the equation of motion. The zeroth order equation is
\be
{\Psi^{(0)}_\pm}''+\frac{f'}{f} { \Psi^{(0)}_\pm }'+\frac{1}{f} \left(\frac{w^2}{f}-Q z ~ C_\pm(0)-\frac{f'}{z}\right)\Psi^{(0)}_\pm =0
\ee
and second order equation is
\be
{ \Psi_\pm^{(2)} }'' + \frac{f'}{f} { \Psi_\pm^{(2)} }'+ \frac{1}{f}\left(\frac{w^2}{f}-Q z ~ C_\pm(0)-\frac{f'}{z}\right)\Psi^{(2)}_\pm -\frac{1+Q z ~C_\pm'(0)}{f} \Psi^{(0)}_\pm =0 .
\ee
Numerically, ${\cal{G}}_\pm^{(0)}, {\cal{G}}_\pm^{(2)}$ are computed as
\be
{\cal{G}}_\pm^{(0)} = \lim_{\epsilon \rightarrow 0}\frac{{\phi_1^{(0)}}'(\epsilon)}{\phi^{(0)}_1(\epsilon)} , \quad {\cal{G}}_\pm^{(2)} = \lim_{\epsilon \rightarrow 0}\frac{{\phi_1^{(2)}}'(\epsilon)}{\phi^{(2)}_1(\epsilon)}.
\ee

\section{Negative index of refraction}
At the beginning, let us first review the electric permittivity
$\epsilon(\omega)$ and the magnetic permeability $\mu(\omega)$ and
their dependence on the Green functions. At the leading order of the
electromagnetic coupling, the optical properties of the medium can
be described by the linear response to an external electromagnetic
field. In an isotropic medium with spatial dispersion, the electric
permittivity $\epsilon(\omega,k)$ and the magnetic permeability
$\mu(\omega,k)$ depend on both the frequency and the wave vector
$k$. The refractive index for the transverse modes has the form \be
n^2(\omega,k)=\epsilon(\omega,k)\mu(\omega,k) \ee Without the
dissipation, $\epsilon(\omega,k)$ and $\mu(\omega,k)$ are real.
However, in case of that the medium is dissipative,
$\epsilon(\omega,k)$ and $\mu(\omega,k)$ will have imaginary part.
So that the refractive index $n$ is a complex quantity. The real
part of $n$ is the refraction index while the imaginary part encodes
the information of dissipation. The refraction can be negative if
$Re(\epsilon)$ and $Re(\mu)$ are not simultaneously negative.  For
the dissipative case, the negative refractive index is equivalent to
require
\be
n_{DL}=|\epsilon(\omega)| Re(\mu(\omega))+ |\mu(\omega)|
Re(\epsilon(\omega))<0.
\ee
In the Laudau-Lifshitz approach to electrodynamics of continuous
media \cite{17, 18, 19}, the transverse part of the dielectric
tensor is determined by the transverse retarded Green function as
follows \cite{19,20}
\be
\epsilon_{T}=1-\frac{4\pi e^2}{\omega^2}G_{T}(\omega,k),
\ee
where $e^2$ is the 3-dimensional electromagnetic
coupling and $G_{T}(\omega,k)$ is the transverse part of the Green
function.

\begin{figure} \
\begin{center}
    \includegraphics[angle=0, width=0.43 \textwidth]{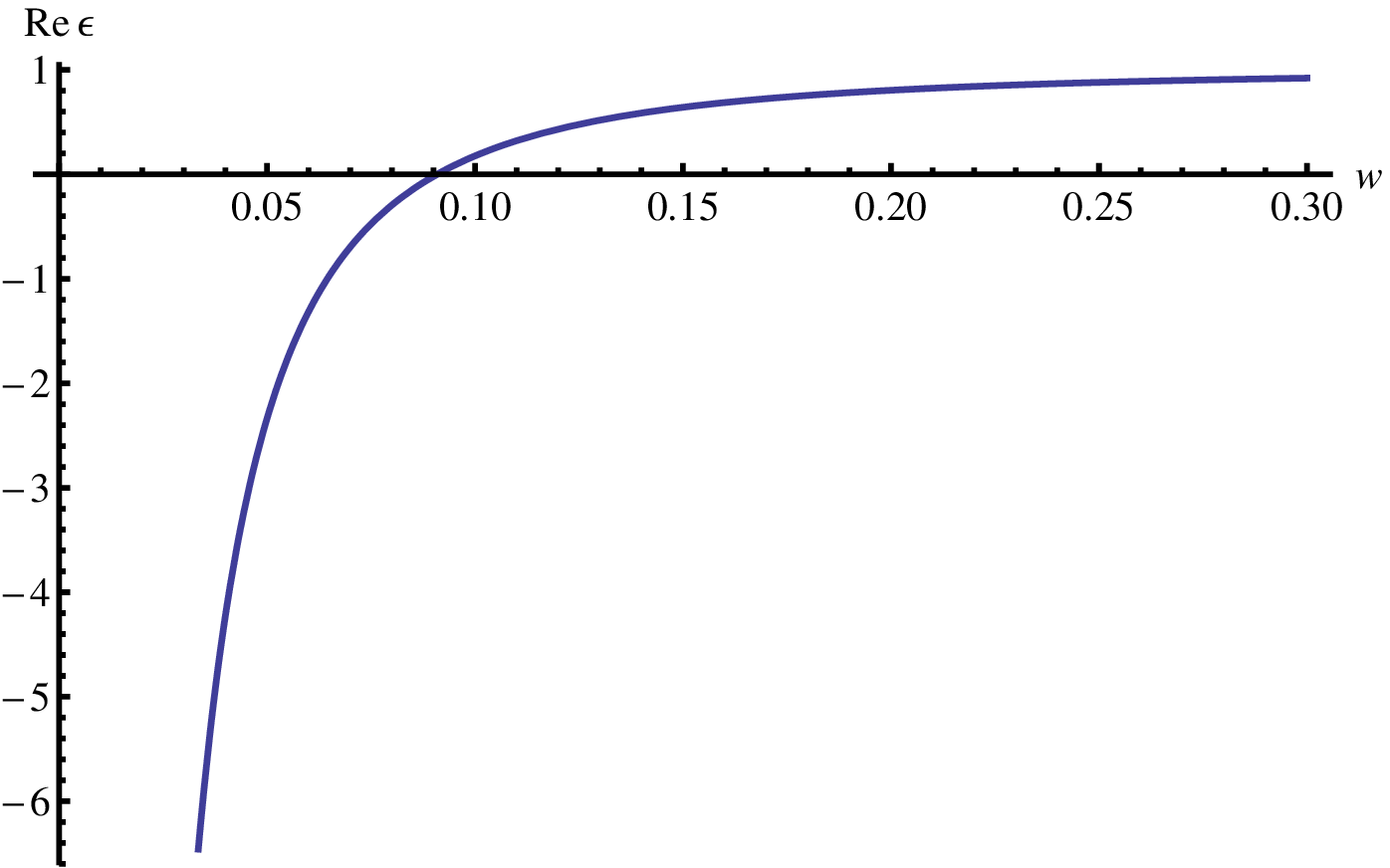}
    \includegraphics[angle=0, width=0.43 \textwidth]{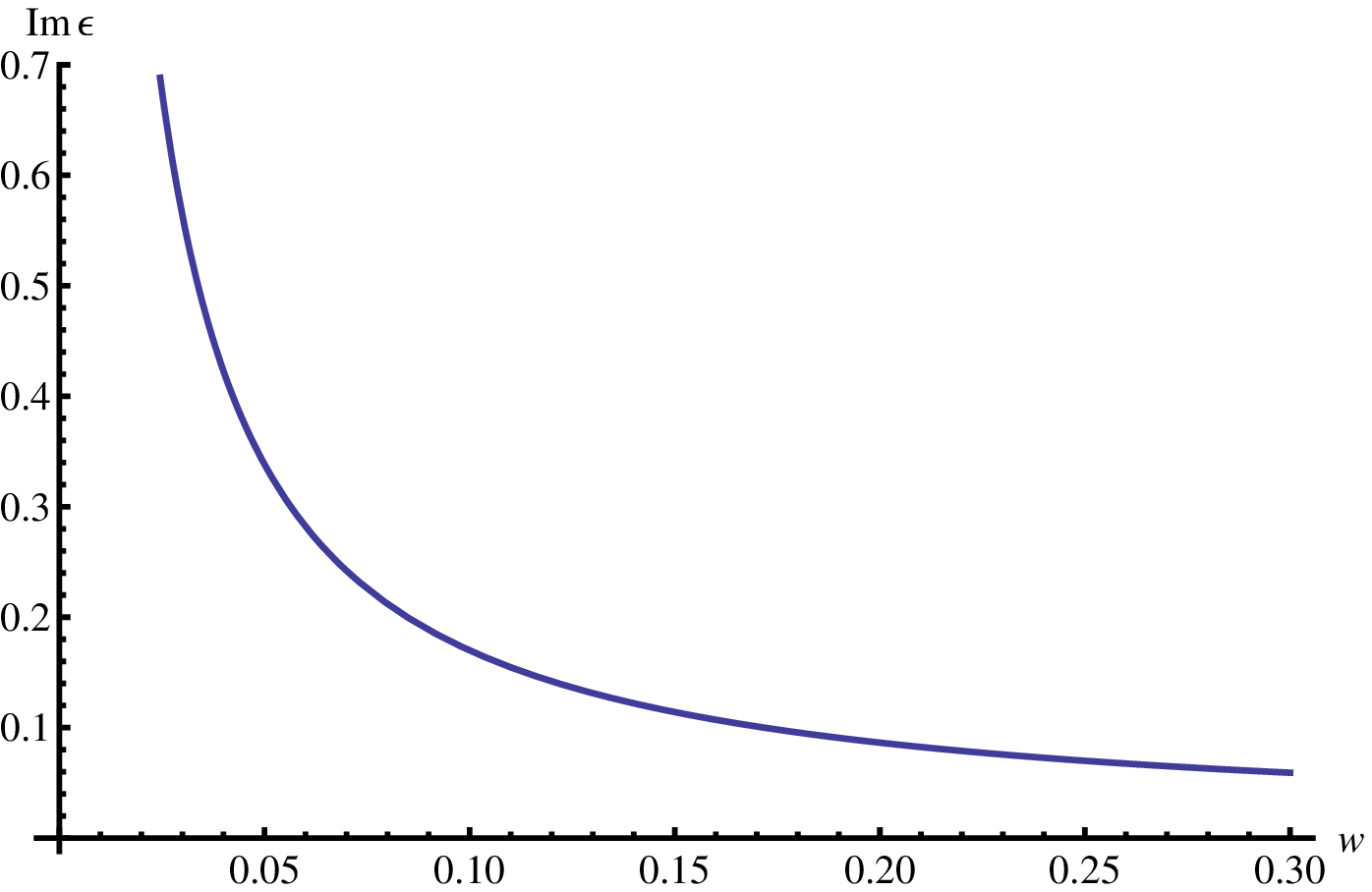}
  \caption{Real(left) and imaginary(right) part of transverse electric permittivity, $\epsilon$(w) with $e$ =0.5 and Q=0.5 .}
\end{center}
\end{figure}
\begin{figure} \
\begin{center}
    \includegraphics[angle=0, width=0.43 \textwidth]{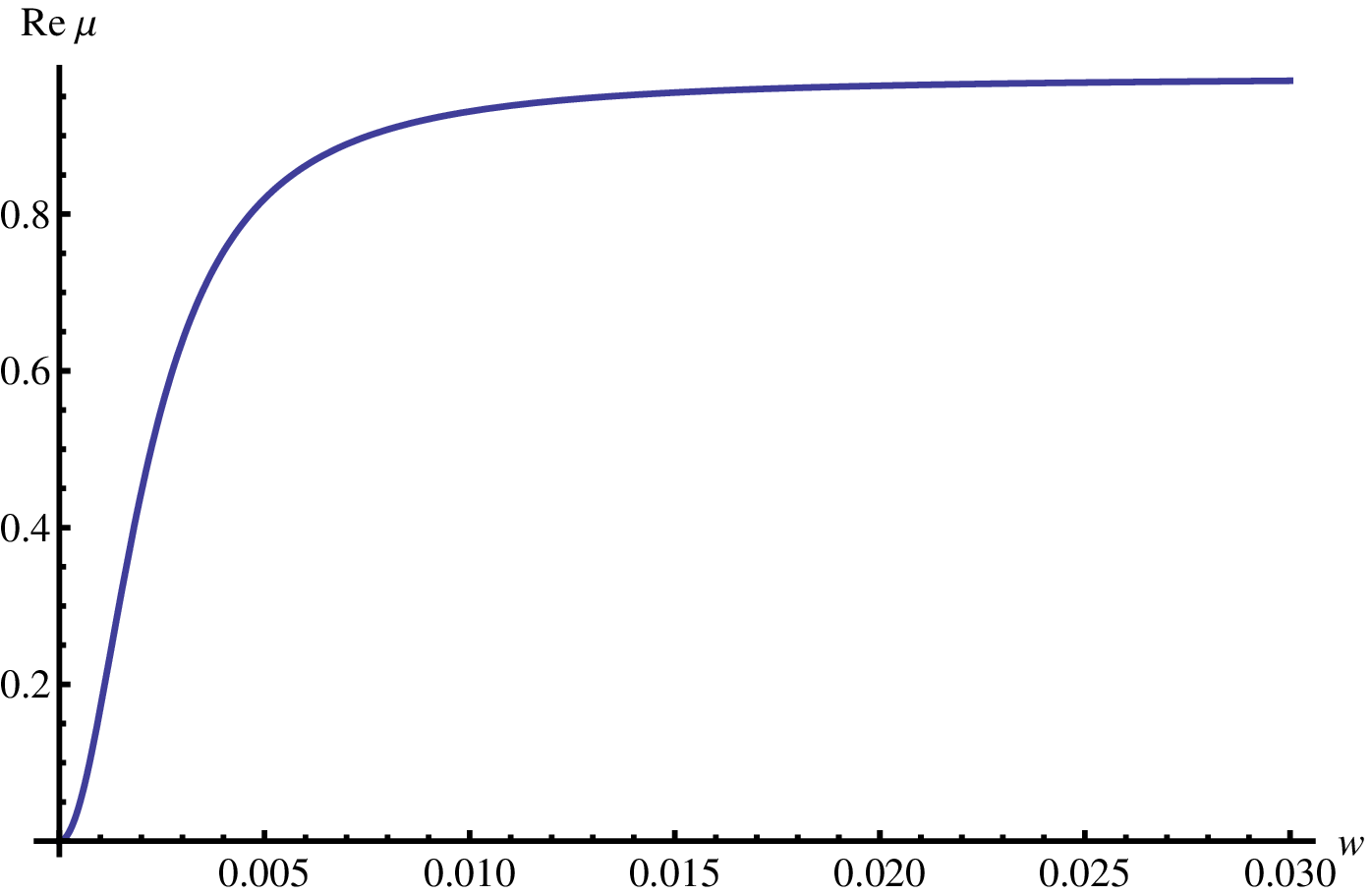}
    \includegraphics[angle=0, width=0.43 \textwidth]{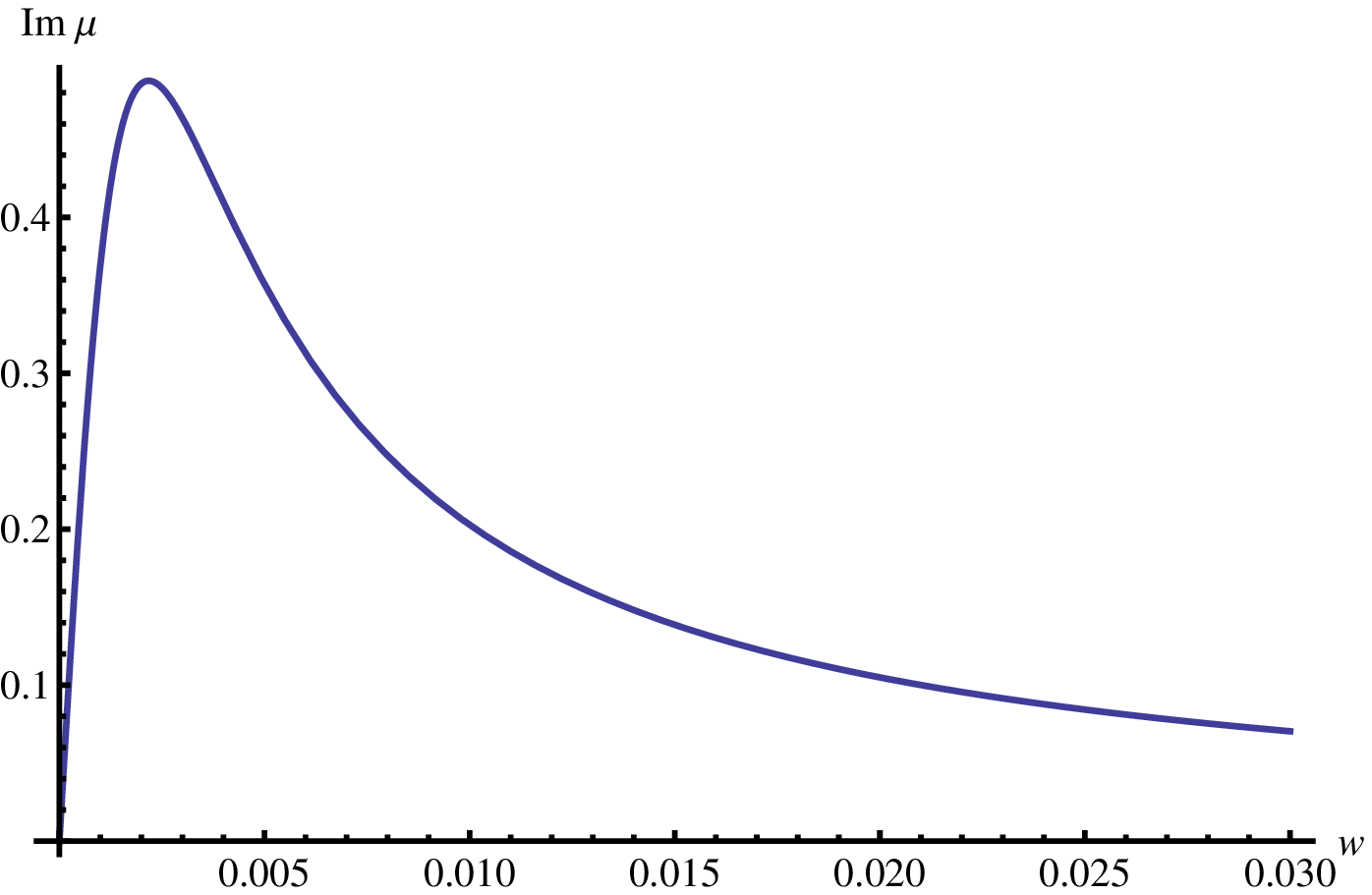}
  \caption{Real(left) and imaginary(right) part of transverse magnetic permeability $\mu$(w) with $e$ =0.5 and Q=0.5.}
\end{center}
\end{figure}

By expanding the Green function as
$G_{T}(\omega,k)=G^{(0)}_{T}(\omega)+k^2 G^{(2)}_{T}(\omega)$,
\be \label{G2N}
{\cal{G}}^{(2)} = \frac{9+21 Q^2-Q^4+3 Q^6}{9 (1+Q^2)^3}-\frac{4 i Q^2}{9 (1+Q^2)^2 w}
\ee
the electric permittivity and effective magnetic permeability are given by
\ba
&&\epsilon(\omega)=1-\frac{4\pi e^2}{\omega^2}G^{(0)}_{T}(\omega,k)=1+i\frac{4\pi e^2 \sigma(\omega)}{\omega} \label{31} \\
&&\mu(\omega)=\frac{1}{1+4\pi e^2 G^{(2)}_{T}(\omega)}. \label{32}
\ea
we use the eq. (\ref{condKubof}) for the expression of $\epsilon$(w).
For sufficiently small value of e, the eq. ($\ref{32}$) can be expanded as
\be
\mu(\omega) = 1- 4\pi e^2 G^{(2)}_{T}(\omega) + \mathcal{O}(e^4)
\ee
From the above equations, we can see that at low enough frequencies, $n_{DL}$ can be negative. From eq.
(\ref{G2N}), (\ref{31}) and (\ref{32}), in hydrodynamic limit,  we find that
\ba
\epsilon(\omega)&=&1+\frac{4\pi e^2 l^2 \alpha}{(1+Q^2)\kappa^2_{4}}
\bigg(\frac{i}{\omega}\frac{(3-Q^2)^2}{9\alpha(1+Q^2)}-\frac{1}{\omega^2}\frac{4Q^2}{3(1+Q^2)}\bigg), \\
\mu(\omega) &=&\left(1+\frac{4\pi e^2 l^2 \alpha}{(1+Q^2)\kappa^2_{4}} \left(\frac{9+21 Q^2-Q^4+3 Q^6}{9 (1+Q^2)^3}-\frac{4 i Q^2}{9 (1+Q^2)^2 w}\right)\right)^{-1} \label{mu1} \\
& = & 1+ \frac{i}{\omega}\frac{16\pi e^2 Q^2}{9(1+Q^2)\alpha}+\mathcal{O}(e^4), \quad \mbox{for} ~~ e^2/\omega \ll 1. \label{mu2}
\ea
The last equality can not be valid for small enough frequency,
causing the difference of our figure for the permeability $\mu$ from that in the ref. \cite{amariti}.
Our result suggest that at static limit, the medium exhibiting the  character of the super conductivity, i.e, the zero permeability. 

It is clear that the value of
$n_{DL}=|\epsilon(\omega)| Re(\mu(\omega))+ |\mu(\omega)|
Re(\epsilon(\omega))$ can be negative at the low frequency regime,
since the real part of $\epsilon(\omega)$ acquires negative value,
while the imaginary part of $\mu(\omega)$ is positive.
\begin{figure} \
\begin{center}
    \includegraphics[angle=0, width=0.5 \textwidth]{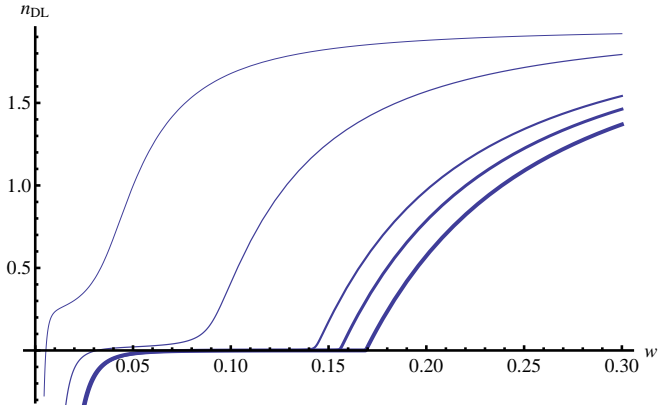}
  \caption{Refractive index $n_{DL}$ for Q=0.2, 0.5, 1, 1.2, 1.5 (from left to right). }
\end{center}
\end{figure}

\section{Conclusion}
In this paper, we  worked out the momentum dependent
hydrodynamic analysis for the vector modes of   charged AdS$_4$   black hole.
  We also calculated  the spectral function of the dual
field theory and as an application, we calculated the permittivity and permeability
and found that for low frequency regime, the index of refraction is found to be
negative, which  support the  claim  made in ref.\cite{amariti} for AdS$_5$.

It will be very interesting if one can extend our analysis to the sound modes, which will enable
us to calculate the longitudinal permittivity.
It will be also interesting  to workout the the case involving the magnetic charge.

 \vspace*{10mm} \noindent
 {\large{\bf Acknowledgments}}\\
\vspace{1mm}
SJS want to thank Shanghai University for the kind hospitality during his visit for this work. Also KHJ appreciate A. Amariti for the useful discussion. The work of XHG  was partly supported by NSFC, China
(No. 10947116 and No. 11005072),  Shanghai Rising-Star Program and
SRF for ROCS SEM. KHJ and SJS was also supported by Mid-career Researcher Program through NRF grant (No. 2010-0008456 ), and by the WCU project (R33-2008-000- 10087-0) and also  by NRF grant through the CQUeST with grant number 2005-0049409. And the work of KHJ is supported by the Seoul Fellowship.

\renewcommand{\theequation}{A.\arabic{equation}}

\setcounter{equation}{0} \setcounter{footnote}{0}

\end{document}